\documentstyle[aps,manuscript]{revtex}
\begin{document}
\draft
\title{Tendency towards Maximum Complexity in a Non-equilibrium
Isolated System}
\author{Xavier Calbet}
\address{Instituto de Astrof{\'\i}sica de Canarias\\
V{\'\i}a L\'actea, s/n,\\
 E-38200 La Laguna, Tenerife, Spain\\
E-mail: xca@ll.iac.es
}
\author{Ricardo L\'opez-Ruiz}
\address{
Departamento de F{\'\i}sica Te\'orica\\
Facultad de Ciencias, Edificio A,
Universidad de Zaragoza,\\
E-50009 Zaragoza,
 Spain\\
E-mail: rilopez@posta.unizar.es}

\date{\today}
\maketitle
\begin{abstract}
The time evolution equations of a simplified isolated
ideal gas, the ``tetrahedral'' gas, are derived.
The dynamical behavior of  LMC (L\'opez, Mancini, Calbet)
complexity
is studied in this system.
In general, it is shown that
the complexity remains within the bounds
of minimum and maximum complexity.
We find that there are certain
restrictions when the isolated ``tetrahedral'' gas evolves
towards equilibrium.
In addition to the well-known increase in entropy,
the quantity called disequilibrium decreases
monotonically with time.
Furthermore, the trajectories of the system
in phase space approach the maximum complexity path
as it evolves toward equilibrium.

\end{abstract}

\section{Introduction}
\label{sec:intro}

Several definitions of complexity, in the general sense of the
term, have been
presented in the literature.
These can be classified according to their
calculation procedure into two broad and loosely
defined groups.

One of these groups is based on computational science and
consists of all definitions based on
algorithms or automata to derive the complexity.
Examples are the logical depth \cite{bennet1986},
the $\epsilon$-machine complexity
\cite{crutchfield1989},
 and
algorithmic complexity \cite{chaitin1966}.
These definitions have been shown to be very useful in describing symbolic
dynamics of chaotic maps, but they have the disadvantage of being
very difficult to calculate.

Another broad group consists of those complexities
based on the measure of entropy or entropy rate.
Among these, we may cite the metric or K--S entropy
rate \cite{kolmogorov1958,sinai1959},
the thermodynamic depth \cite{lloyd1988}, the effective measure complexity
\cite{grassberger1986},
and the simple measure for complexity \cite{shiner1999}.
These definitions have also been very useful in describing
symbolic dynamical maps, the latter  having been applied
to a non-equilibrium Fermi gas
\cite{landsberg1998}.
They suffer the disadvantage of either being  very difficult
to calculate or having a simple relation to the regular
entropy.

New definition types of complexity have recently been
introduced. These are based on quantities that can be calculated
directly from the distribution function describing the system.
One of these is based on ``meta-statistics''
\cite{atmanspacher1997} and the other
on the notion of ``disequilibrium'' \cite{lopez1995}.
This latter definition will be referred to hereafter as the
LMC complexity.
These definitions, together with the simple measure for
complexity \cite{shiner1999}
described above,
have the great advantage of allowing easy calculations
within the context of kinetic theory and of permitting their evaluation
in a natural way in terms of statistical mechanics.

The disequilibrium-based complexity is
easy to calculate and shows some interesting properties
\cite{lopez1995}, but suffers
from the main drawback of not being very well behaved
as the system size increases, or equivalently, as the distribution
function becomes continuous
\cite{feldman1998}.
Feldman and Crutchfield tried to solve this problem
by defining another equivalent term for disequilibrium, but
ended up with a complexity that was a trivial function of the
entropy.

Whether these definitions of complexity are useful
in non-equilibrium thermodynamics will depend
on how they behave as a function of time.
There is a general belief that, although
the second law of thermodynamics
requires average entropy (or disorder) to increase,
this does not in any way forbid local order from arising
\cite{gell-mann1995}.
The clearest example is seen with life, which can continue
to exist and grow in an isolated system for as long as
internal resources last.
In other words, in an isolated system the entropy must increase,
but it should be possible, under certain circumstances, for
the complexity to increase.

In this paper we will examine how LMC complexity
evolves with time in an isolated system and
we will show that it indeed has some interesting properties.
The disequilibrium-based
complexity defined in Ref. \cite{lopez1995} actually tends to be maximal
as the entropy increases in a Boltzmann integro--differential
equation for a simplified gas.

In Sec.\ \ref{sec:def} LMC complexity
definition is reviewed.
We proceed to calculate the distributions
which maximize and minimize the complexity
and its asymptotic behavior,
and also introduce the basic
concepts underlying the time evolution of LMC
complexity in Sec.\ \ref{sec:time}.
Later, in Sec.\ \ref{sec:tetra},
by means of numerical computations following
a restricted version of
the
Boltzmann equation, we apply this to a special
system, which we shall term ``tetrahedral gas''.
Finally, in Sec.\ \ref{sec:conclu},
the results and possible future lines of investigation
 are discussed, together with their possible applications.
In the Appendix is given an
analytical and a numerical demonstrations of the results of
the numerical calculations
for the tetrahedral gas.

\section{Definition of Complexity}
\label{sec:def}

The definition of LMC complexity, $C$, is

\begin{equation}
\label{eq:defc}
C = D \cdot H,
\end{equation}

where $D$ is the disequilibrium term and $H$ is the entropy.

We assume that the system can be in one of $N$ possible
accessible states, $i$. The probability of the system being in state $i$
will be given by the discrete distribution function, $f_i$.
The system is defined such that, if isolated,
it will reach equilibrium, with all the states having equal probability,
$f_{\rm e}$.

These definitions imply that
all values of $f_i$ are positive,

\begin{equation}
f_i \ge 0,
\label{eq:pos}
\end{equation}
and that a normalization, $I$,
must hold such that

\begin{equation}
I \equiv \sum_{i=1}^N f_i = 1,
\label{eq:norm}
\end{equation}
and the equilibrium distribution function is,

\begin{equation}
f_{\rm e} = \frac{1}{N}.
\end{equation}

The definition of disequilibrium, $D$, is given as a distance
to the probability in equilibrium, $f_{\rm e}$:

\begin{equation}
D \equiv \sum_{i=1}^N (f_i - f_{\rm e})^2.
\label{eq:defd}
\end{equation}

The normalized entropy, $H$, is defined as

\begin{equation}
H \equiv - \frac{1}{\ln N} \sum_{i=1}^N f_i \ln f_i.
\label{eq:defh}
\end{equation}

Note that since $0 \leq H \leq 1$ and
$0 \leq D \leq (N-1)/N$,
the complexity, $C$, is normalized
($0 \leq C \leq 1$).

\section{Complexity versus time}
\label{sec:time}

\subsection{Complexity versus entropy}

We are interested in an isolated system
with an initial arbitrary discrete distribution, and which
evolves toward equilibrium, in which it reaches an
equiprobability distribution.

To study the time evolution of the complexity,
a diagram of $C$ versus time, $t$, can be used. But, as
we know,
the second law of thermodynamics states that
the entropy grows monotonically with time; that is,

\begin{equation}
\frac{d H}{d t} \geq 0.
\end{equation}

This implies that an equivalent way to study the
time evolution of the complexity
can be obtained by plotting $C$ versus $H$.
In this way, the entropy substitutes the
time axis, since the former increases monotonically
with the latter.
The conversion from $C$ vs. $H$ to $C$ vs. $t$ diagrams
is achieved by stretching or shrinking
the entropy axis according to its time evolution.
This method is a key point in all this discussion.
Note that, in any case,
the relationship of $H$ versus $t$ will, in general, not be a simple
one \cite{latora1999}.

\subsection{Maximum and minimum complexity}

When an isolated system evolves with time, the
complexity cannot have any possible value
in a $C$ versus $H$ map, but it
must stay within certain bounds, $C_{\rm max}$ and $C_{\rm min}$.
These are the maximum and minimum values of $C$
for a given $H$.
Since $C = D \cdot H$, finding the extrema of $C$ for constant $H$
is equivalent to finding the extrema of $D$.

There are two restrictions on $D$:
the normalization, $I$, and the fixed value of the
entropy, $H$.
To find these extrema undetermined Lagrange multipliers
are used. Differentiating
Eqs.\ (\ref{eq:norm}),
(\ref{eq:defd}) and (\ref{eq:defh})
we obtain,

\begin{eqnarray}
\frac{\partial D}{\partial f_j} & = & 2(f_j - f_{\rm e}),\\
\frac{\partial I}{\partial f_j} & = & 1,\\
\frac{\partial H}{\partial f_j} & = & -\frac{1}{\ln N}\left(\ln f_j + 1\right).
\end{eqnarray}

Defining $\lambda_1$ and $\lambda_2$ as the Lagrange
multipliers, we get:

\begin{equation}
2(f_j - f_{\rm e}) + \lambda_1 + \lambda_2(\ln f_j + 1)/\ln N = 0.
\end{equation}

Two
new parameters, $\alpha$ and $\beta$,
which are a
linear combinations of the Lagrange multipliers are defined:

\begin{equation}
f_j + \alpha \ln f_j + \beta = 0,
\label{eq:maxmin}
\end{equation}
where the solutions of this equation, $f_j$,
are the values that minimize or
maximize the disequilibrium.

In the maximum complexity case
there are two solutions, $f_j$, to Eq.\ (\ref{eq:maxmin})
which are shown in Table \ref{tab:maximum}.
One of these solutions,
$f_{\rm max}$, is given by

\begin{equation}
\label{eq:defhmax}
H = - \frac{1}{\ln N} \left[ f_{\rm max} \ln f_{\rm max} + ( 1 - f_{\rm max} ) \ln
\left( \frac{1 - f_{\rm max}}{N - 1} \right) \right],
\end{equation}
and the other solution by $(1 - f_{\rm max})/(N - 1)$.

The maximum disequilibrium, $D_{\rm max}$, for a fixed $H$ is

\begin{equation}
\label{eq:defdmax}
D_{\rm max} = (f_{\rm max} - f_{\rm e})^2 +
      (N - 1)\left(\frac{1 - f_{\rm max}}{N - 1} - f_{\rm e}\right)^2,
\end{equation}
and thus, the maximum complexity, which depends
only on $H$, is

\begin{equation}
\label{eq:defcmax}
C_{\rm max}(H) = D_{\rm max} \cdot H.
\end{equation}

Note that the behavior  of the maximum value of complexity
versus $\ln N$ has been studied
in Ref. \cite{anteneodo1996}.

Equivalently, the values, $f_j$, that give a minimum complexity
are shown in Table \ref{tab:minimum}. One of the solutions,
$f_{\rm min}$, is given by

\begin{equation}
H = - \frac{1}{\ln N} \left[ f_{\rm min} \ln f_{\rm min} + ( 1 - f_{\rm min} ) \ln
     \left( \frac{1 - f_{\rm min}}{N - n - 1} \right) \right],
\end{equation}
where $n$ is the number of states with $f_j = 0$
and takes a value in the range $n = 0, 1,\ \ldots\ ,N - 2$.

The resulting
minimum disequilibrium, $D_{\rm min}$, for a given $H$ is,

\begin{equation}
D_{\rm min} = (f_{\rm min} - f_{\rm e})^2 + (N - n - 1)
        \left(\frac{1-f_{\rm min}}{N - n - 1} -
f_{\rm e}\right)^2 + n f_{\rm e}^2.
\end{equation}

Note that in this case $f_j = 0$  is an additional
hidden solution that stems from the positive restriction
in the $f_i$ values, Eq.\ (\ref{eq:pos}).
To obtain these solutions explicitly
we can define $x_i$ such that

\begin{equation}
f_i \equiv {x_i}^2.
\end{equation}

These $x_i$ values do not have the restriction imposed by
Eq.\ (\ref{eq:pos}) and can take a positive or negative value.
If we repeat the Lagrange multiplier method with these
new variables a new solution arises:
$x_j = 0$, or equivalently, $f_j = 0$.

The resulting minimum complexity, which again only
depends on $H$, is

\begin{equation}
\label{defcmin}
C_{\rm min}(H)= D_{\rm min} \cdot H.
\end{equation}

As an example, the maximum and minimum of complexity,
$C_{\rm max}$ and $C_{\rm min}$,
are plotted
as a function of the entropy, $H$, in Fig.~\ref{fig:c_minmax}
for $N=4$.
In Fig.~\ref{fig:d_minmax} the maximum and minimum disequilibrium,
$D_{\rm max}$ and $D_{\rm min}$, versus $H$
are also shown.

\subsection{Minimum ``envelope''}

As shown in Fig.~\ref{fig:d_minmax}
the minimum disequilibrium function is piecewise
defined, having several points where its derivative
is discontinuous. Each of these function pieces
corresponds to a different value of
$n$ (Table \ref{tab:minimum}).
In some circumstances it might be helpful
to work with the ``envelope'' of the minimum disequilibrium
function. The function, $D_{\rm minenv}$, that traverses
all the discontinuous derivative points in the $D_{\rm min}$
versus $H$ plot is

\begin{equation}
D_{\rm minenv} = e^{- H \ln N} - \frac{1}{N},
\end{equation}
and is also shown in Figure~\ref{fig:d_minmax}.

\subsection{Asymptotic behavior of the complexity
as {\boldmath $N \rightarrow \infty$}}

When $N$ tends toward infinity the probability, $f_{\rm max}$,
of the dominant state has a linear dependence
with the entropy,

\begin{equation}
\lim_{N \rightarrow \infty} f_{\rm max} = 1 - H,
\end{equation}
and thus the maximum disequilibrium scales as

\begin{equation}
\lim_{N \rightarrow \infty} D_{\rm max} = (1-H)^2.
\end{equation}

The maximum complexity tends to

\begin{equation}
\label{eq:cmaxlim}
\lim_{N \rightarrow \infty} C_{\rm max} = H \cdot (1-H)^2.
\end{equation}

The limit of the minimum disequilibrium and complexity
vanishes,

\begin{equation}
\lim_{N \rightarrow \infty} D_{\rm minenv} = 0,
\end{equation}
and thus

\begin{equation}
\label{eq:cminlim}
\lim_{N \rightarrow \infty} C_{\rm min} = 0.
\end{equation}

In general, in the limit $N \rightarrow \infty$,
the complexity is not a trivial function of the entropy,
in the sense that for a given $H$ there exists
a range of complexities between $0$ and $C_{\rm max}$
(Eqs. [\ref{eq:cminlim}] and [\ref{eq:cmaxlim}]).

In particular, in this asymptotic limit,
the maximum of $C_{\rm max}$ is found when
$H=1/3$, or equivalently $f_{\rm max}=2/3$,
which gives a maximum of the maximum complexity
of $C_{\rm max}=4/27$.
This was numerically calculated by Anteneodo and
Plastino in Ref. \cite{anteneodo1996}.

\section{An example: the tetrahedral gas}
\label{sec:tetra}

\subsection{The tetrahedral gas}

We present a simplified example of an ideal gas: the tetrahedral
gas.  This system is generated by a
simplification of the Boltzmann integro--differential equation
of an ideal gas. We are interested in studying the disequilibrium time
evolution.

The Boltzmann integro--differential equation
of an ideal gas with no external forces and no spatial gradients is

\begin{equation}
\label{eq:boltzman}
\frac{\partial f({\bf v};t)}{\partial t} =
        \int d^{3} {\bf v}_{*} \int d \Omega_{\rm c.m.}
        \sigma( {\bf v}_{*} - {\bf v} \rightarrow
        {\bf v}'_{*} - {\bf v}' )
        | {\bf v}_{*} -
        {\bf v} | \left[ f({\bf v}'_{*};t)
        f({\bf v}';t) - f({\bf v}_{*};t) f({\bf v};t)
        \right],
\end{equation}
where $\sigma$ represents the cross section of a collision
between two particles with initial velocities ${\bf v}$ and
${\bf v}_{*}$ and after the collision with velocities
${\bf v}'$ and ${\bf v}'_{*}$; and $\Omega_{\rm c.m.}$ are
all the possible dispersion angles of the collision as
seen from its center of mass.

In the tetrahedral gas,
the particles can travel only in four directions
in three-dimensional space and all have
the same absolute velocity. These directions
are the ones given by joining the center of a tetrahedron
with its corners.
The directions can be easily viewed by recalling
the directions given by a methane molecule, or equivalently,
by a caltrop, which is a device with four metal points
so arranged that when any three are on the ground the fourth
projects upward as a hazard to the hooves of horses or to
pneumatic tires (see Fig.~\ref{fig:tetra}).

By definition, the angle that one direction forms with any
other is the same.
It can be shown that the angles between different directions,
$\alpha$, satisfy the relationship $\cos \alpha = - 1/3$,
which gives $\alpha=109.47^{\circ}$. The plane formed by any
two directions is perpendicular to the plane formed by the
remaining two directions.

We assume that the cross-section, $\sigma$, is different from
zero only when the angle between the velocities of the colliding
particles is $109.47^{\circ}$. It is also assumed that this collision
makes the two particles leave in the remaining two directions,
thus again forming an angle of $109.47^{\circ}$.
A consequence of these restrictions is that the
modulus of the velocity is always the same no matter how many collisions
a particle has undergone and they always stay within the directions
of the vertices of the tetrahedron. Furthermore, this type of gas
does not break any law of physics and is perfectly valid, although
hypothetical.

We label the four directions originating from
the center of the caltrop with numbers, ${\bf 1},
{\bf 2}, {\bf 3}, {\bf 4}$
(see Fig.~\ref{fig:tetra}).
The velocity components with the same direction but opposite
sense, or equivalently,
directed toward the center of the caltrop,
are labeled with negative numbers ${\bf -1}, {\bf -2},
{\bf  -3}, {\bf -4}$.

In order to formulate the Boltzmann equation for the tetrahedral gas,
and because all directions are equivalent,
we need only study the different collisions that a particle with
one fixed direction can undergo.
In particular if we take a particle with direction ${\bf -1}$
the result of the collision with another particle
with direction ${\bf -2}$ are the same two
particles traveling in directions ${\bf 3}$ and ${\bf 4}$;
that is,

\[({\bf -1}, {\bf -2}) \rightarrow ( {\bf 3}, {\bf 4}). \]

With this in mind the last bracket of Eq. (\ref{eq:boltzman})
is,

\[f_{3} f_{4} - f_{-1} f_{-2}, \]
where $f_{i}$ denotes the probability of finding
a particle in direction ${\bf i}$.
Note that
the dependence on velocity, ${\bf v}$,
of the continuous velocity distribution
function, $f({\bf v};t)$, of Eq. (\ref{eq:boltzman})
is in our case contained in the discrete subindex,
$i$, of the distribution function $f_i$.

We can proceed in the same manner with the other
remaining collisions,

\begin{eqnarray*}
({\bf -1}, {\bf -3}) &\rightarrow ( {\bf 2}, {\bf 4}), \\
({\bf -1}, {\bf -4}) &\rightarrow ( {\bf 2}, {\bf 3}).
\end{eqnarray*}

When a particle with direction ${\bf -1}$ collides with a particle
with direction ${\bf 2}$, they do not form an angle
of $109.47^\circ$; i.e., they
do not collide, they just pass by each other.
This is a consequence of the previous assumption for the
tetrahedral gas,
which establishes a null cross section for angles different
from $109.47^\circ$.
The same can be said
for collisions $({\bf -1}, {\bf 3})$, $({\bf -1}, {\bf 4})$, and
$({\bf -1}, {\bf 1})$.
All these results are summarized in Table \ref{tab:col}.

Taking all this into account,
Eq. (\ref{eq:boltzman}) for
direction ${\bf -1}$ is reduced to a discrete sum,

\begin{equation}
\frac{d f_{-1}}{d t} =
                       ( f_{3} f_{4} - f_{-1} f_{-2} ) +
                       ( f_{2} f_{4} - f_{-1} f_{-3} ) +
                       ( f_{2} f_{3} - f_{-1} f_{-4} ),
\end{equation}
where all other factors have been set to unity for simplicity.

The seven remaining equations are:

\begin{eqnarray}
\frac{d f_{-2}}{d t} = ( f_{3} f_{4} - f_{-1} f_{-2} ) +
                                ( f_{1} f_{4} - f_{-2} f_{-3} ) +
                                ( f_{1} f_{3} - f_{-2} f_{-4} ), \nonumber \\
\frac{d f_{-3}}{d t} = ( f_{2} f_{4} - f_{-3} f_{-1} ) +
                                ( f_{4} f_{1} - f_{-3} f_{-2} ) +
                                ( f_{1} f_{2} - f_{-3} f_{-4} ), \nonumber \\
\frac{d f_{-4}}{d t} = ( f_{2} f_{3} - f_{-4} f_{-1} ) +
                                ( f_{3} f_{1} - f_{-4} f_{-2} ) +
                                ( f_{1} f_{2} - f_{-4} f_{-3} ), \nonumber \\
\frac{d f_{1}}{d t} = ( f_{-3} f_{-4} - f_{1} f_{2} ) +
                                ( f_{-2} f_{-4} - f_{1} f_{3} ) +
                                ( f_{-2} f_{-3} - f_{1} f_{4} ), \nonumber \\
\frac{d f_{2}}{d t} = ( f_{-3} f_{-4} - f_{1} f_{2} ) +
                                ( f_{-1} f_{-4} - f_{2} f_{3} ) +
                                ( f_{-2} f_{-3} - f_{2} f_{4} ), \nonumber \\
\frac{d f_{3}}{d t} = ( f_{-2} f_{-4} - f_{3} f_{1} ) +
                                ( f_{-1} f_{-4} - f_{3} f_{2} ) +
                                ( f_{-1} f_{-2} - f_{3} f_{4} ), \nonumber \\
\frac{d f_{4}}{d t} = ( f_{-2} f_{-3} - f_{4} f_{1} ) +
                                ( f_{-1} f_{-3} - f_{4} f_{2} ) +
                                ( f_{-1} f_{-2} - f_{3} f_{4} ).
\end{eqnarray}

If we now make $f_{i} = f_{-i} ( i=1,2,3,4 )$ initially,
this property is conserved in time.
The final four equations defining the evolution of the
system are:

\begin{eqnarray}
\label{eq:tetra}
\frac{d f_{1}}{d t} = ( f_{3} f_{4} - f_{1} f_{2} ) +
                                ( f_{2} f_{4} - f_{1} f_{3} ) +
                                ( f_{2} f_{3} - f_{1} f_{4} ), \nonumber \\
\frac{d f_{2}}{d t} = ( f_{3} f_{4} - f_{1} f_{2} ) +
                                ( f_{1} f_{4} - f_{2} f_{3} ) +
                                ( f_{1} f_{3} - f_{2} f_{4} ), \nonumber \\
\frac{d f_{3}}{d t} = ( f_{2} f_{4} - f_{3} f_{1} ) +
                                ( f_{1} f_{4} - f_{3} f_{2} ) +
                                ( f_{1} f_{2} - f_{3} f_{4} ), \nonumber \\
\frac{d f_{4}}{d t} = ( f_{2} f_{3} - f_{4} f_{1} ) +
                                ( f_{1} f_{3} - f_{4} f_{2} ) +
                                ( f_{1} f_{2} - f_{3} f_{4} ).
\end{eqnarray}

Note that the ideal gas has been reduced to
the tetrahedral gas, which is a four-dimensional
dynamical system.
The velocity distribution function, $f_i$,
corresponds to the probability distribution function
of Sec.\ \ref{sec:def} with $N=4$ accessible states.

\subsection{Evolution of the tetrahedral gas with time}
\label{sec:evol}

The tetrahedral gas (Eqs. [\ref{eq:tetra}])
reaches equilibrium
when $f_{i} = 1/N$ for $i=1,2,3,4$ and $N=4$.
This stationary state, $d f_i / d t = 0$, represents the equiprobability
towards which the system evolves in time.
This is consistent with
the previous definition of disequilibrium, Eq. (\ref{eq:defd}),
in which we assumed that equilibrium was reached
at equiprobability, where $D=0$.

As the isolated system evolves it gets closer and closer to equilibrium.
In this sense, one may intuitively think that the disequilibrium
will decrease with time.
In fact, it can be shown
that, as the system approaches  equilibrium, $D$
tends to zero monotonically with time:

\begin{equation}
\label{eq:dtlez}
\frac{d D}{d t} \le 0.
\end{equation}

The analytical demonstration of this inequality
for the tetrahedral gas is shown in Appendix
\ref{sec:dt}.

There are even more restrictions on the evolution
of this system. It would be expected that
the system approaches equilibrium, $D=0$, by
following the most direct path.
To verify this, numerical simulations for several initial conditions
have been undertaken.
In all of these we observe the additional restriction that $D$
approaches $D_{\rm max}$ on its way to $D=0$. In fact it appears
as an exponential decay of $D$ towards $D_{\rm max}$ in a $D$ versus $H$ plot.
As an example, two of these are shown in Fig.~\ref{fig:evold}, where
Fig.~\ref{fig:evold}(a) shows
a really strong tendency towards $D_{\rm max}$.
Contrary to intuition, among all the possible
paths that the system can follow toward equilibrium,
it chooses those closest to $D_{\rm max}$ in particular.

We can also observe this effect in a
complexity, $C$, versus $H$ plot.
This is shown for the same two initial conditions
as the previous figure in Figure~\ref{fig:evolc}.

This additional restriction
to the evolution of the system is better viewed by
plotting the difference $C_{\rm max} - C$ versus $H$.
In all the cases analyzed(see two in Fig.~\ref{fig:evolcmax})
the following condition is observed:

\begin{equation}
\frac{d (C_{\rm max} - C)}{d t} \le 0.
\end{equation}

This has been verified numerically and is
illustrated in Figure~\ref{fig:cmaxmct}, where
this time derivative, which always remains negative, is shown as a function
of $H$ for a grid of uniformly spaced distribution functions,
$(f_1, f_2, f_3, f_4)$, satisfying the normalization condition,
Eq.\ (\ref{eq:norm}).
Two system trajectories are also shown for illustrative
purposes. The numerical method used to plot this function is explained in
Appendix \ref{sec:cmaxmct}.

\subsection{Maximum complexity path as an attractive trajectory}

As shown in Table \ref{tab:maximum},
a collection of maximum complexity
distributions for $N=4$ can take the form

\begin{eqnarray}
\label{eq:coll}
f_1 &=& f_{\rm max} \nonumber \\
f_i &=& \frac{1-f_{\rm max}}{3}, i=2,3,4
\end{eqnarray}
where $f_{\rm max}$ runs from $1/N$ (equiprobability distribution)
to $1$ (``crystal'' distribution). The complexity of
this collection
of distributions covers all possible values of $C_{\rm max}$.

There is actually a time evolution of the
tetrahedral gas,
or trajectory of the system, formed
by this collection of distributions.
Inserting Eqs. (\ref{eq:coll})
in the evolution
Eqs. (\ref{eq:tetra}),
it is found that all equations are compatible
with each other and the dynamical equations are reduced
to the relation,

\begin{equation}
\frac{d f_{\rm max}}{d t} = \frac{1}{3} ( 4 f_{\rm max}^2 - 5 f_{\rm max} + 1).
\end{equation}
This trajectory is denoted as the {\it maximum complexity path}.

Note that the equiprobability
or equilibrium, $f_{\rm max} = 1/4$, is a  stable fixed point and
the  maximum disequilibrium ``crystal'' distribution,
$f_{\rm max}=1$, is an unstable fixed point.
Thus
the maximum complexity path is a heteroclinic
connection between the ``crystal'' and equiprobability distributions.

The maximum complexity path is locally attractive.
Let us assume, for instance, the following perturbed trajectory

\begin{eqnarray}
f_1 &=& f_{\rm max}, \nonumber \\
f_2 &=& \frac{1-f_{\rm max}}{3}, \nonumber \\
f_3 &=& \frac{1-f_{\rm max}}{3} + \delta, \nonumber \\
f_4 &=& \frac{1-f_{\rm max}}{3} - \delta,
\end{eqnarray}
whose evolution according to Eqs. (\ref{eq:tetra}) gives the
exponential decay of the perturbation, $\delta$:

\begin{equation}
\frac{d \delta}{d t} \sim - \left( \frac{4 f_{\rm max} + 2}{3} \right) \delta,
\end{equation}
showing the attractive nature of these trajectories.

\section{Conclusion}
\label{sec:conclu}

The time evolution of the LMC complexity, $C$,
has been studied for a simplified model of an isolated
ideal gas: the tetrahedral gas.
In general, the dynamical behavior of this quantity is bounded
between two extremum curves, $C_{\rm max}$ and $C_{\rm min}$,
when observed in
a $C$ versus $H$ phase space.
These complexity bounds have been derived.

For the isolated tetrahedral gas
two constraints on its dynamics are found.
The first, which is analytically demonstrated, is
that the disequilibrium, $D$, decreases monotonically with
time until it reaches the value $D=0$ for the equilibrium
state.
The second is that the maximum complexity paths, $C_{\rm max}$,
are attractive in phase space.
In other words,
the complexity of the system tends to equilibrium
always approaching those paths. We verify this  numerically;
that is, the time derivative of the difference between
$C_{\rm max}$ and $C$ is negative.

Fig.~\ref{fig:suma} summarizes the dynamical
behavior of the tetrahedral gas.
The different trajectories starting  with arbitrary
initial conditions, which represent systems out of
equilibrium, evolve towards equilibrium approaching
the maximum complexity path.

Whether these properties are useful in real
physical systems will depend on further work
on this subject. More can be said about the macroscopical
nature of the disequilibrium when this work is extended
to more general systems, such as to the
ideal gas following the complete Boltzmann integro--differential
equation. Another feature, which could result useful,
would be to approximate the evolution of a real physical
system trajectory to its maximum complexity path.
Note that in general, for a real system, the calculation of the maximum
complexity path will not be an easy task.

\appendix
\section{Proof of the monotonically decrease in disequilibrium
with time}
\label{sec:dt}

We now present the analytical proof of the inequality
shown in Eq. (\ref{eq:dtlez}) from Sec.\ \ref{sec:evol},

\begin{equation}
\label{eq:dtlezapp}
\frac{d D}{d t} \le 0.
\end{equation}

The time derivative of the disequilibrium, $D$,
for the isolated
tetrahedral gas, is explicitly given by,

\begin{equation}
\frac{d D}{d t} = \frac{d}{d t} \sum_{i} (f_i - f_{\rm e})^{2} =
\sum_{i} 2 f_{i} \frac{d f_{i}}{d t} - \sum_{i} 2 f_{\rm e} \frac{d f_i}{d t}.
\end{equation}

Using the normalization Eq. (\ref{eq:norm}) we find that

\begin{equation}
\label{eq:dnorm}
\sum_{i} \frac{d f_{i}}{d t} = 0,
\end{equation}
and we are left with

\begin{equation}
\frac{d D}{d t} = 2 \sum_{i} f_i \frac{d f_i}{d t}.
\end{equation}

If we substitute Boltzmann Eq. (\ref{eq:tetra}) in the
previous one, we obtain:

\begin{eqnarray}
\label{eq:dt}
F \equiv \frac{1}{2} \frac{d D}{d t} =
&+ f_1 f_3 f_4 - f_1^2 f_2 + f_1 f_2 f_4 -
             f_1^2 f_3 + f_1 f_2 f_3 - f_1^2 f_4 \nonumber \\
&+ f_2 f_3 f_4 - f_1 f_2^2 + f_1 f_2 f_4 -
             f_3 f_2^2 + f_1 f_2 f_3 - f_2^2 f_4 \nonumber \\
&+ f_1 f_3 f_4 - f_3^2 f_2 + f_2 f_3 f_4 -
             f_1 f_3^2 + f_1 f_2 f_3 - f_3^2 f_4 \nonumber \\
&+ f_1 f_2 f_4 - f_3 f_4^2 + f_1 f_3 f_4 -
             f_2 f_4^2 + f_2 f_3 f_4 - f_1 f_4^2,
\end{eqnarray}
where the new variable $F$ has been defined.

We now split this function $F$ into two different terms:

\begin{eqnarray}
G \equiv & - f_1^2 f_2 - f_2^2 f_1 - f_3^2 f_1 - f_3^2 f_2 \nonumber \\
         & - f_4^2 f_1 - f_4^2 f_2 \nonumber \\
         & + f_1 f_3 f_4 + 2 f_1 f_2 f_3 \nonumber \\
         & + 2 f_1 f_2 f_4 + f_2 f_3 f_4
\end{eqnarray}
and

\begin{eqnarray}
K \equiv & - f_3^2 f_4 - f_4^2 f_3 \nonumber \\
         & - f_1^2 f_3 - f_1^2 f_4 \nonumber \\
         & - f_2^2 f_3 - f_2^2 f_4 \nonumber \\
         & + f_1 f_2 f_3 + 2 f_1 f_3 f_4 \nonumber \\
         & + 2 f_2 f_3 f_4 + f_1 f_2 f_4.
\end{eqnarray}
Then

\begin{equation}
F = G + K.
\end{equation}

Note that $G$ and $K$ are symmetrical functions,
in the sense that
we can transform one into the other by changing $(f_1, f_2, f_3, f_4)$
by $(f_3, f_4, f_1, f_2)$, respectively.

To prove that $G \le 0$,
the following variable change is performed:

\begin{eqnarray}
f_1 & = & y_1, \nonumber \\
f_2 & = & y_2, \nonumber \\
f_3 & = & y_3 + y_4, \nonumber \\
f_4 & = & y_3 - y_4.
\end{eqnarray}

The positivity of the distributions functions,
Eq. (\ref{eq:pos}), implies that, $y_1 \ge 0$ and $y_2 \ge 0$.

The function $G$ in the new variables reads,

\begin{equation}
G = - y_1^2 y_2 - y_2^2 y_1 - y_3^2 y_1 - 3 y_4^2 y_1 - y_3^2 y_2
- 3 y_4^2 y_2 + 4 y_1 y_2 y_3.
\end{equation}

Regrouping terms, $G$ can be expressed as

\begin{equation}
G = - y_2 ( y_1 - y_3 )^2 - y_1 ( y_2 - y_3 )^2
- 3 y_4^2 y_1 - 3 y_4^2 y_1.
\end{equation}

Since $y_1$ and $y_2$ are both positive
and the squared quantities are also positive,
we conclude that

\begin{equation}
G \le 0.
\end{equation}

The same inequality can be demonstrated for $K$ due
to its symmetry with $G$, thus proving the assumption
from Eq. (\ref{eq:dtlezapp}).

\section{Method of calculating the time derivative
of the maximum complexity minus the complexity}
\label{sec:cmaxmct}

To calculate the quantity
\begin{equation}
\frac{d (C_{\rm max} - C)}{d t}
\end{equation}
from some given values
of the distribution functions, $(f_1, f_2, f_3, f_4)$,
we derive the expression of maximum complexity,
Eq.\ (\ref{eq:defcmax}), minus the definition of complexity,
Eq.\ (\ref{eq:defc}), obtaining,

\begin{equation}
\frac{d (C_{\rm max} - C)}{d t} = \frac{d D_{\rm max}}{d t} H +
D_{\rm max} \frac{d H}{d t} - \frac{d D}{d t} H - D \frac{d H}{d t}.
\end{equation}

Let us now examine each of these terms beginning from the end.
The time derivative of the entropy is calculated
by differentiating its definition, Eq.\ (\ref{eq:defh}):

\begin{equation}
\frac{d H}{d t} = - \frac{1}{\ln N} \sum_{i}
\left[ \frac{d f_i}{d t} \ln f_i + \frac{d f_i}{d t} \right].
\end{equation}
But recalling Eq.\ (\ref{eq:dnorm}) we can simplify this to

\begin{equation}
\frac{d H}{d t} = - \frac{1}{\ln N} \sum_{i}
\left[ \frac{d f_i}{d t} \ln f_i \right].
\end{equation}
This last term can be easily calculated using the evolution
Eqs.\ (\ref{eq:tetra}).

The terms $D$ and $H$ can be readily calculated using
their definitions from Eqs.\ (\ref{eq:defd}) and (\ref{eq:defh}).

The time derivative of the disequilibrium has already been
expressed in Appendix \ref{sec:dt} with Eq.\ (\ref{eq:dt}).

The maximum disequilibrium is calculated with
the previous found value of $H$ by inverting Eq.\ (\ref{eq:defhmax})
and obtaining $f_{\rm max}$. After this, the expression defining
the maximum disequilibrium, Eq.\ (\ref{eq:defdmax}), can be used.

Finally the time derivative of the maximum disequilibrium is calculated
as follows:

\begin{equation}
\frac{d D_{\rm max}}{d t} = \frac{d D_{\rm max}}{d f_{\rm max}} \;
\frac{d f_{\rm max}}{d H} \; \frac{d H}{d t}.
\end{equation}

The first two derivatives on the right-hand side can be calculated
analytically with Eqs.\ (\ref{eq:defdmax}) and (\ref{eq:defhmax}).
The numerical value can be found using the previously
calculated value of $f_{\rm max}$.


%
%

\begin{figure}
\caption{Maximum, minimum, and minimum envelope complexity, $C_{\rm max}$,
$C_{\rm min}$, and $C_{\rm minenv}$ respectively, as a function of
the entropy, $H$, for a system with $N=4$ accessible states.}
\label{fig:c_minmax}
\end{figure}

\begin{figure}
\caption{Maximum, minimum, and minimum envelope disequilibrium, $D_{\rm max}$,
$D_{\rm min}$, and $D_{\rm minenv}$ respectively, as a function of
the entropy, $H$, for a system with $N=4$ accessible states.}
\label{fig:d_minmax}
\end{figure}

\begin{figure}
\caption{The four possible directions of the velocities of the
tetrahedral gas in space. Positive senses are defined
as emerging from the center point and with integer numbers,
$1, 2, 3, 4$.}
\label{fig:tetra}
\end{figure}

\begin{figure}
\caption{Time evolution of the system in $(H,D)$ phase space
for two different initial conditions at time $t=0$:
(a) $(f_1, f_2, f_3, f_4) = (0.8, 0.2, 0, 0)$
and (b) $(f_1, f_2, f_3, f_4) = (0.5, 0.5, 0, 0)$.
The maximum and minimum disequilibrium are shown by
dashed lines.}
\label{fig:evold}
\end{figure}

\begin{figure}
\caption{Time evolution of the system in $(H,C)$ phase space
for two different initial conditions at time $t=0$:
(a) $(f_1, f_2, f_3, f_4) = (0.8, 0.2, 0, 0)$
and (b) $(f_1, f_2, f_3, f_4) = (0.5, 0.5, 0, 0)$.
The maximum and minimum complexity are shown by
dashed lines.}
\label{fig:evolc}
\end{figure}

\begin{figure}
\caption{Time evolution of the system in $(H,C_{\rm max}-C)$ phase space
for two different initial conditions at time $t=0$:
(a) $(f_1, f_2, f_3, f_4) = (0.8, 0.2, 0, 0)$
and (b) $(f_1, f_2, f_3, f_4) = (0.5, 0.5, 0, 0)$.
The values $C_{\rm max} - C_{\rm min}$ are shown by
dashed lines.}
\label{fig:evolcmax}
\end{figure}

\begin{figure}
\caption{Numerical verification of $d (C_{\rm max} - C)/d t \le 0$.
This time derivative is shown as a function of $H$.
A grid of uniformly spaced, $\Delta f_{i} = 0.5$,
distribution functions, $(f_1, f_2, f_3, f_4)$,
satisfying the normalization condition
Eq.\ (\ref{eq:norm}), have been used.
Two system trajectories for initial
conditions, $t=0$, $(f_1, f_2, f_3, f_4) = (0.8, 0.2, 0, 0)$
and $(f_1, f_2, f_3, f_4) = (0.5, 0.5, 0, 0)$ are also shown for illustrative
purposes.
It can be seen how the above-mentioned time derivative always remains
negative.}
\label{fig:cmaxmct}
\end{figure}

\begin{figure}
\caption{Summary of this paper. The time evolution of the system for
three different initial conditions,
$t=0$, $(f_1, f_2, f_3, f_4) = (0.8, 0.2, 0, 0)$,
 $(f_1, f_2, f_3, f_4) = (0.5, 0.5, 0, 0)$, and the maximum
complexity path are shown.
The minimum complexity is shown by dashed lines.
It can be seen how the system tends to approach the maximum complexity
path as it evolves in time toward equilibrium.}
\label{fig:suma}
\end{figure}

%
%

\narrowtext
\begin{table}
\caption{Probability values, $f_j$, that give a maximum
of disequilibrium, $D_{\rm max}$, for a given $H$}
\label{tab:maximum}
\begin{tabular}{ccc}
Number of states with
    & $\displaystyle f_j$ & Range of $\displaystyle f_j$\\
with $\displaystyle f_j$ & & \\
\tableline
$\displaystyle 1$         & $f_{\rm max}$
         & $\displaystyle \frac{1}{N}\ \ldots\ 1$\\
$\displaystyle N - 1$     & $\displaystyle \frac{1 - f_{\rm max}}{N - 1}$ &
         $\displaystyle 0\ \ldots\ \frac{1}{N}$\\
\end{tabular}
\end{table}

\begin{table}
\caption{Probability values, $f_j$, that give a minimum
of disequilibrium, $D_{\rm min}$, for a given $H$}
\label{tab:minimum}
\begin{tabular}{ccc}
Number of states
         & $\displaystyle f_j$ & Range of $\displaystyle f_j$\\
with $\displaystyle f_j$ \tablenotemark[1] & & \\
\tableline
$\displaystyle n$         & $\displaystyle 0$
         & $\displaystyle 0$\\
$\displaystyle 1$     & $\displaystyle f_{\rm min}$ &
         $\displaystyle 0\ \ldots\ \frac{1}{N - n}$\\
$\displaystyle N - n - 1$     &
         $\displaystyle \frac{1 - f_{\rm min}}{N - n - 1}$ &
         $\displaystyle \frac{1}{N - n}\ \ldots\ \frac{1}{N - n - 1}$\\
\end{tabular}
\tablenotetext[1]{$n$ can have the values $0, 1,\ \ldots\, N-2$}
\end{table}

\begin{table}
\caption{Cross sections, $\sigma$, for a particle
in direction ${\bf -1}$ colliding
with particles in the other remaining
directions of the tetrahedral gas}
\label{tab:col}
\begin{tabular}{cc}
Collision       &     Cross section \\
                &     $\sigma$      \\
\tableline
$({\bf -1}, {\bf -2}) \rightarrow ( {\bf 3}, {\bf 4})$  &  $1$ \\
$({\bf -1}, {\bf -3}) \rightarrow ( {\bf 2}, {\bf 4})$  &  $1$ \\
$({\bf -1}, {\bf -4}) \rightarrow ( {\bf 2}, {\bf 3})$  &  $1$ \\
Other collisions                                        & $0$ \\
\end{tabular}
\end{table}

\end{document}